\documentclass[12pt]{article}

\usepackage{feynarts}
\usepackage{epsfig}

\makeatletter

\def\paragraph{\@startsection{paragraph}{4}{\z@}{+2.00ex plus
 +1ex minus +.2ex}{1.5ex plus .2ex}{\it\normalsize}}

\def\section{\@startsection {section}{1}{\z@}{+3.0ex plus +1ex minus
  +.2ex}{2.3ex plus .2ex}{\normalsize\bf\boldmath}}
\def\subsection{\@startsection{subsection}{2}{\z@}{+2.5ex plus +1ex
minus +.2ex}{1.5ex plus .2ex}{\normalsize\bf\boldmath}}
\def\subsubsection{\@startsection{subsubsection}{3}{\z@}{+3.25ex plus
 +1ex minus +.2ex}{1.5ex plus .2ex}{\normalsize\it}}

\expandafter\ifx\csname mathrm\endcsname\relax\def\mathrm#1{{\rm #1}}\fi

\newcounter{saveeqn}

\@addtoreset{equation}{section}

\newcount\@tempcntc
\def\@citex[#1]#2{\if@filesw\immediate\write\@auxout{\string\citation{#2}}\fi
  \@tempcnta\z@\@tempcntb\m@ne\def\@citea{}\@cite{\@for\@citeb:=#2\do
    {\@ifundefined
       {b@\@citeb}{\@citeo\@tempcntb\m@ne\@citea
        \def\@citea{,\penalty\@m\ }{\bf ?}\@warning
       {Citation `\@citeb' on page \thepage \space undefined}}%
    {\setbox\z@\hbox{\global\@tempcntc0\csname
b@\@citeb\endcsname\relax}%
     \ifnum\@tempcntc=\z@ \@citeo\@tempcntb\m@ne
       \@citea\def\@citea{,\penalty\@m}
       \hbox{\csname b@\@citeb\endcsname}%
     \else
      \advance\@tempcntb\@ne
      \ifnum\@tempcntb=\@tempcntc
      \else\advance\@tempcntb\m@ne\@citeo
      \@tempcnta\@tempcntc\@tempcntb\@tempcntc\fi\fi}}\@citeo}{#1}}

\def\@citeo{\ifnum\@tempcnta>\@tempcntb\else\@citea
  \def\@citea{,\penalty\@m}%
  \ifnum\@tempcnta=\@tempcntb\the\@tempcnta\else
   {\advance\@tempcnta\@ne\ifnum\@tempcnta=\@tempcntb \else
\def\@citea{--}\fi
    \advance\@tempcnta\m@ne\the\@tempcnta\@citea\the\@tempcntb}\fi\fi}

\def\asymp#1%
{\mathrel{\raisebox{-.4em}{$\widetilde{\scriptstyle #1}$}}}

\def\Nequal#1%
{\mathrel{\raisebox{-.5em}{$\stackrel{=}{\scriptstyle\rm#1}$}}}
\newcommand{\dsl}[1]{\not \hspace{-0.7mm}#1}
\def\dsl{\mathpalette\make@slash}
\def\make@slash#1#2{\setbox\z@\hbox{$#1#2$}%
  \hbox to 0pt{\hss$#1/$\hss\kern-\wd0}\box0}

\def\beq{\begin{equation}}
\def\eeq{\end{equation}}
\def\beqar{\begin{eqnarray}}
\def\eeqar{\end{eqnarray}}

\def\solid{\raise.9mm\hbox{\protect\rule{1.1cm}{.2mm}}}
\def\dash{\raise.9mm\hbox{\protect\rule{2mm}{.2mm}}\hspace*{1mm}}
\def\dot{\rlap{$\cdot$}\hspace*{2mm}}

\def\Re{\mathop{\mathrm{Re}}\nolimits}
\def\draftdate{\relax}
\def\mda{\relax}
\def\mua{\relax}
\def\mla{\relax}
\def\draft{
\def\thtystars{******************************}
\def\sixtystars{\thtystars\thtystars}
\typeout{}
\typeout{\sixtystars**}
\typeout{* Draft mode!
         For final version remove \protect\draft\space in source file *}
\typeout{\sixtystars**}
\typeout{}
\def\draftdate{\today}
\def\mua{\marginpar[\boldmath\hfil$\uparrow$]%
                   {\boldmath$\uparrow$\hfil}%
                    \typeout{marginpar: $\uparrow$}\ignorespaces}
\def\mda{\marginpar[\boldmath\hfil$\downarrow$]%
                   {\boldmath$\downarrow$\hfil}%
                    \typeout{marginpar: $\downarrow$}\ignorespaces}
\def\mla{\marginpar[\boldmath\hfil$\rightarrow$]%
                   {\boldmath$\leftarrow $\hfil}%
                    \typeout{marginpar: $\leftrightarrow$}\ignorespaces}
\def\Mua{\marginpar[\boldmath\hfil$\Uparrow$]%
                   {\boldmath$\Uparrow$\hfil}%
                    \typeout{marginpar: $\uparrow$}\ignorespaces}
\def\Mda{\marginpar[\boldmath\hfil$\Downarrow$]%
                   {\boldmath$\Downarrow$\hfil}%
                    \typeout{marginpar: $\downarrow$}\ignorespaces}
\def\Mla{\marginpar[\boldmath\hfil$\Rightarrow$]%
                   {\boldmath$\Leftarrow $\hfil}%
                    \typeout{marginpar: $\leftrightarrow$}\ignorespaces}
\overfullrule 5pt
\oddsidemargin -15mm
\marginparwidth 29mm
}

\def\stars{\strut\leaders\hbox{*}\hfill\strut}
\def\starline{\hfil\strut\hfil\hbox to \textwidth {\stars}\hfil}

\begin{document}
\thispagestyle{empty}
\def\thefootnote{\fnsymbol{footnote}}
\setcounter{footnote}{1}
\null
\strut\hfill PSI-PR-04-05\\
\strut\hfill hep-ph/0403251
\vfill
\begin{center}
{\Large \bf\boldmath{Electron to Muon Conversion in Low-Energy 
Electron-Nucleus Scattering}
\par} \vskip 2.5em
\vspace{1cm}

Kai-Peer~O.~Diener \\[1cm]
Paul Scherrer Institut\\
CH-5232 Villigen PSI, Switzerland \\[0.5cm]
\end{center}\par
\vskip 2cm  
We present an estimate of the electron to muon conversion cross section in
fixed-target elastic electron scattering. 
The matrix element \mbox{$\langle \mu | j_\mathrm{em}^\mu(0) | e \rangle $} 
is calculated analytically in two scenarios introducing suitable approximations.
We consider on the one hand side the case of three light Dirac neutrinos with CKM-type leptonic 
mixing and on the other hand a typical see-saw scenario.
We evaluate the coulombic contribution to the scattering cross section
in the limit of vanishing energy transfer to the nucleus and, thus, obtain
a realistic estimate for the total conversion cross section.
Although we find that in the see-saw scenario the cross section can be enhanced 
by as much as twenty orders of magnitude in comparison to the Dirac case, 
it is still not experimentally accessible.

\par
\vskip 3cm
\noindent
March 2004
\null
\setcounter{page}{0}
\clearpage
\def\thefootnote{\arabic{footnote}}
\setcounter{footnote}{0}

\clearpage

\section{Introduction}
\label{sec:intro}

In recent years the physics of neutrino oscillations 
and lepton flavour violation (LFV)
has been a very active field of fundamental research.
Neutrino oscillations could be demonstrated in
several experiments, see e.g.\ Refs.~\cite{Fukuda:1998mi}, 
and massive neutrinos have become part of the Standard Model 
of particle physics (SM).

As neutrinos have very small masses, neutrino mass eigenstates
cannot be identified by direct measurements of their four-momenta.
Thus, in a typical oscillation experiment, neutrino flavour 
is identified indirectly through the mass of the 
charged lepton associated with charged-current production or 
detection of the neutrino.
A distinction is made between disappearance and appearance 
experiments depending on whether the number of 
neutrinos of predefined flavour 
is expected to decrease or increase between production 
and detection, respectively.
Other experiments look for flavour transitions of 
charged leptons either in decays, like \mbox{$\mu \to e \gamma$},
or in boundstates of a nucleus and a captured muon, 
usually referred to as $\mu$-$e$ conversion.
Considerable theoretical and experimental effort has been 
dedicated to $\mu$-$e$ conversion, see e.g.\ Refs.~\cite{Feinberg:1959ui,Vergados:1985pq},
but conversion experiments, so far, have not been able
to test or complement the results on LFV 
found in neutrino oscillation experiments.

In this paper we are instead concerned with the question 
whether LFV could be observed directly in fixed-target
elastic electron scattering, i.e.\ in the process
\mbox{$e + \mathrm{N} \to \mu + \mathrm{N}$}, 
at energies low compared
to the mass of the $W$-boson or a heavy nucleus.
Clearly, the reduced timescale of the interaction
makes it less likely to observe LFV
in a scattering process than in a classical conversion experiment.
The great advantages of a scattering experiment, on the other hand,
lie in the simplicity of the experimental setup and the theoretical
description, which 
relies on no nuclear properties other than electric charge 
and allows us to get by without addressing the demanding
problems of the QED muon-nucleus boundstate.

The following section 
will be concerned with the 
calculation of the matrix element 
\mbox{$\langle \mu | j_\mathrm{em}^\mu(0) | e \rangle $}.
We give compact results for the electric and magnetic form
factors, first in the limit of small neutrino masses and 
external momenta, then for heavy Majorana neutrinos.
LFV is parameterised in the context of the SM by including 
a leptonic CKM-type matrix, customarily called the 
MNS~\cite{Maki:mu} matrix. 
Majorana neutrinos are introduced through the standard see-saw
mechanism, leading to one light, almost purely left-handed and 
one (or more) heavy, mostly right-handed self-conjugate neutrino per
SM generation.
In section~\ref{sec:Xsec} we derive the cross section  for 
electron to muon conversion 
through coulombic interaction with a heavy nucleus at 
(laboratory frame) electron energies 
\mbox{$E_e \approx 2 m_\mu$}.
Section~\ref{sec:res} contains a discussion of our results as
well as our conclusions.

\section{The Amplitude \mbox{$ e \to \mu \gamma^\star $}}
\label{sec:amp}

\begin{figure}
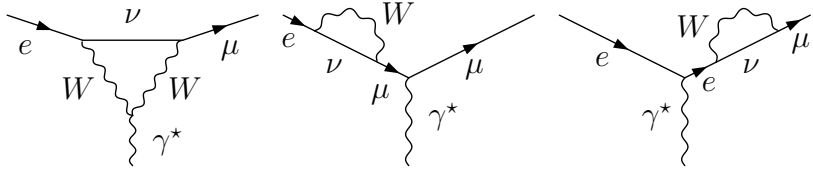

\centerline{
\unitlength=1bp%
\begin{feynartspicture}(532,104)(3,1)
\FADiagram{}
\FAProp(0.,15.)(6.,13.)(0.,){/Straight}{1}
\FALabel(2.,13.)[tr]{$e$}
\FALabel(10.7,15.5)[tr]{$\nu$}
\FALabel(7.,10.)[tr]{$W$}
\FALabel(15.5,10.)[tr]{$W$}
\FAProp(20.,15.)(14.,13.)(0.,){/Straight}{-1}
\FALabel(18.5,13.)[tr]{$\mu$}
\FAProp(10.,3.)(10.,7.)(0.,){/Sine}{0}
\FAProp(10.,7.)(6.,13.)(0.,){/Sine}{0}
\FAProp(10.,7.)(14.,13.)(0.,){/Sine}{0}
\FAProp(6.,13.)(14.,13.)(0.,){/Straight}{0}
\FALabel(14.,6.)[tr]{$\gamma^\star$}
\FADiagram{}
\FAProp(0.,15.)(2.5,13.75)(0.,){/Straight}{1}
\FAProp(7.5,11.25)(10.,10.)(0.,){/Straight}{1}
\FAProp(2.5,13.75)(7.5,11.25)(0.,){/Straight}{0}
\FAProp(2.5,13.75)(7.5,11.25)(-1.,){/Sine}{0}
\FALabel(1.,13.5)[tr]{$e$}
\FALabel(4.75,11.5)[tr]{$\nu$}
\FALabel(10.5,16.)[tr]{$W$}
\FALabel(8.5,9.5)[tr]{$\mu$}
\FAProp(20.,15.)(10.,10.)(0.,){/Straight}{-1}
\FALabel(16.,11.5)[tr]{$\mu$}
\FAProp(10.,3.)(10.,10.)(0.,){/Sine}{0}
\FALabel(14.,8.)[tr]{$\gamma^\star$}
\FADiagram{}
\FAProp(0.,15.)(10.,10.)(0.,){/Straight}{1}
\FALabel(4.,12.)[tr]{$e$}
\FAProp(20.,15.)(17.5,13.75)(0.,){/Straight}{-1}
\FAProp(12.5,11.25)(10.,10.)(0.,){/Straight}{-1}
\FAProp(12.5,11.25)(17.5,13.75)(0.,){/Straight}{0}
\FAProp(12.5,11.25)(17.5,13.75)(-1.,){/Sine}{0}
\FALabel(20.,13.5)[tr]{$\mu$}
\FALabel(16.,11.5)[tr]{$\nu$}
\FALabel(12.,15.)[tr]{$W$}
\FALabel(12.5,10.)[tr]{$e$}
\FAProp(10.,3.)(10.,10.)(0.,){/Sine}{0}
\FALabel(9.,8.)[tr]{$\gamma^\star$}
\end{feynartspicture}
}
\caption{Generic lowest-order diagrams contributing to \mbox{$e \to \mu \gamma^\star$}.
Note that contributions where one or both of the 
$W$-bosons are replaced by their would-be Goldstone counterparts need also be considered. \label{fig:diags}}
\end{figure}

We introduce the notation \mbox{$k_e, k_\mu, q$} 
for the incoming electron and outgoing muon and photon momenta.
Owing to Lorentz symmetry and current conservation that for an on-shell 
electron and muon, the matrix element can be written as~\cite{Feinberg:1959ui}
\beq
  \langle \mu | j_{\mathrm{em}}^\mu (0) | e \rangle =  \sum_{i=1}^4 F_i \mathcal{M}_i^\mu,  
\eeq 
with 
\beqar
   \mathcal{M}_1^\mu &=& \bar u(\vec k_\mu) \left( \gamma^\mu - q^\mu 
	\frac{m_e-m_\mu}{q^2} \right) u(\vec k_e), \\
   \mathcal{M}_2^\mu &=& \bar u(\vec k_\mu) \left( \gamma^\mu  + q^\mu 
	\frac{m_e+m_\mu}{q^2} \right) \gamma^5  u(\vec k_e), \\
   \mathcal{M}_3^\mu &=& \bar u(\vec k_\mu) \left( \gamma^\mu - 
	\frac{k_e^\mu+k_\mu^\mu}{m_e+m_\mu} \right) u(\vec k_e), \\ 
   \mathcal{M}_4^\mu &=& \bar u(\vec k_\mu) \left( \gamma^\mu + 
	\frac{k_e^\mu+k_\mu^\mu}{m_e-m_\mu} \right) \gamma^5 u(\vec k_e). 
\eeqar
The formfactors $F_i$ can be calculated perturbatively and are in general
dimensionless functions of the electron and muon mass, 
the photon virtuality, 
and the masses and couplings of all virtual particles.
We shall confine ourselves to the lowest non-vanishing order in 
perturbation theory, corresponding to the one-loop self-energies
and vertex corrections depicted in 
Fig.~\ref{fig:diags}.
The formfactors are explicitly calculated, first,
in the simplest conceivable extension of the SM allowing
for LFV, with three Dirac neutrinos and a CKM-type, three by three
unitary MNS matrix $V$, then in a see-saw scenario with
one light and one or more heavy Majorana neutrinos per generation.

\subsection{The Case of Three Light Dirac Neutrinos}
\label{sec:Dirac}

In this model, at lowest non-vanishing order, 
the $F_i$ can be factorised in the following way:
\beq
   F_i = \sum_{j=1}^3 e g^2 V_{\nu_j e} V^\star_{\nu_j \mu} 
	\int \frac{\mathrm{d}^D k}{i \pi^2} f_i (m_{\nu_j}^2, m_W^2, q^2, m_e^2, m_\mu^2; k),
\eeq
where $e$ and \mbox{$g=e/(\sqrt{2} s_W)$} denote the electromagnetic and weak coupling constants.
For phenomenological neutrino mass values
the functions $f_i$ contain four very different mass squared
scales spanning over twenty orders of magnitude and
their exact analytical form is, therefore, not suitable for
numerical evaluation.\\
In a first approximation, we expand the functions $f_i$ in 
\mbox{$\epsilon_j = m_{\nu_j}^2/m_W^2$} around \mbox{$\epsilon_j=0$} 
and set the smallest neutrino mass $m_{\nu_1}$ 
to zero. Using the unitarity of the MNS mixing matrix, we obtain
\beq
   F_i = e g^2 \left(
		V_{\nu_2 e} V^\star_{\nu_2 \mu} \epsilon_2  + 
		V_{\nu_3 e} V^\star_{\nu_3 \mu} \epsilon_3  \right)
		\left. \int \frac{\mathrm{d}^D k}{i \pi^2}
	\frac{\partial f_i}{\partial \epsilon_j} \right|_{\epsilon_j = 0} 
		+ \ldots ,
	\label{eq:fexp}
\eeq
where the ellipsis stands for higher orders in $\epsilon_j$, which for all 
practical purposes can safely be neglected.

Because the outer four-momenta are small in comparison to
the $W$-boson mass, we rewrite the propagator denominators containing the loop momentum
$k$, any linear combination of  external momenta $k_\mathrm{ext}$, and the $W$-boson 
mass~\cite{vanderBij:1983bw}:
\beq
      \frac{1}{(k+k_\mathrm{ext})^2-m_W^2} = \frac{1}{k^2-m_W^2} 
		\left( 1 - \frac{2 k k_\mathrm{ext} + k_\mathrm{ext}^2}{(k+k_\mathrm{ext})^2-m_W^2} \right).
	\label{eq:propdec}
\eeq
By iterative application of this identity to the full propagator on the right hand side of 
Eq.~(\ref{eq:propdec}) and truncation of the resulting series, 
we reduce all integrals of Eq.~(\ref{eq:fexp}) to vacuum integrals.
The vacuum integrals can easily be solved, resulting in logarithms of the $W$-boson mass
over the renormalization scale and different powers of $k_\mathrm{ext}^2/m_W^2$,
depending on how many times the decomposition of Eq.~(\ref{eq:propdec}) 
was applied to the full propagators of the loop integrals. 
Due to the ultraviolet finiteness of the total amplitude, all logarithmic 
terms cancel and the final result is a scalar rational function of 
the external four-momenta and $m_W$.
Care must be exercised in the expansion of the loop integral propagators
to obtain a consistent result up to the desired order in $k_\mathrm{ext}^2/m_W^2$, taking
into account the integrals' coefficients.

At the lowest non-trivial order, after at most twofold 
iteration of Eq.~(\ref{eq:propdec}) 
and subsequent truncation of non-vacuum contributions, 
the results of this expansion read
\beqar
\left. \int \frac{\mathrm{d}^D k}{i \pi^2}
	\frac{\partial f_1}{\partial \epsilon_j} \right|_{\epsilon_j = 0} &=& 
	\frac{q^2}{64 \pi^2 m_W^2}, \label{eq:diracff1}\\
\left. \int \frac{\mathrm{d}^D k}{i \pi^2}
	\frac{\partial f_2}{\partial \epsilon_j} \right|_{\epsilon_j = 0} &=& 
	\frac{-q^2}{64 \pi^2 m_W^2}, \\
\left. \int \frac{\mathrm{d}^D k}{i \pi^2}
	\frac{\partial f_3}{\partial \epsilon_j} \right|_{\epsilon_j = 0} &=& 
	\frac{-(m_e+m_\mu)^2}{256 \pi^2 m_W^2}, \\
\left. \int \frac{\mathrm{d}^D k}{i \pi^2}
	\frac{\partial f_4}{\partial \epsilon_j} \right|_{\epsilon_j = 0} &=& 
	\frac{(m_e-m_\mu)^2}{256 \pi^2 m_W^2}. \label{eq:diracff4}
\eeqar
It is a necessary condition that $F_1$ and $F_2$ vanish for \mbox{$q^2 \to 0$},
as the product with the corresponding matrix elements with $q^2$ in the 
denominator would, otherwise, not be well-defined.
We observe, that for vanishing electron mass (and with the aforementioned approximations), 
the total transition 
amplitude attains a particularly simple form:
\beq
	 \langle \mu | j_{\mathrm{em}}^\mu (0) | e \rangle = \frac{e g^2}{32 \pi^2} 
	\left(
		V_{\nu_2 e} V^\star_{\nu_2 \mu} \epsilon_2  + 
		V_{\nu_3 e} V^\star_{\nu_3 \mu} \epsilon_3  \right)
          \left\{ \frac{q^2}{m_W^2} \mathcal{M}_+^\mu
	- \frac{1}{4} \frac{m_\mu^2}{m_W^2} \mathcal{M}_-^\mu \right\},
\eeq
with $\mathcal{M}_+^\mu$ and $\mathcal{M}_-^\mu$ defined as 
\beq
	\mathcal{M}_+^\mu = \bar u(\vec k_\mu) \left( \gamma^\mu + q^\mu 
	\frac{m_\mu}{q^2} \right) L u(\vec k_e), \qquad
	\mathcal{M}_-^\mu = \bar u(\vec k_\mu) \left( \gamma^\mu -
         \frac{k_e^\mu+k_\mu^\mu}{m_\mu} \right) L u(\vec k_e), 
	\label{eq:MpMm}
\eeq
and \mbox{$L = (1-\gamma^5)/2$}, the left-handed  chirality projection operator.

\subsection{The Formfactors in a Typical See-Saw Scenario}
\label{sec:Majorana}

After considering the minimal version of LFV with three light 
Dirac neutrinos in exact 
analogy to flavour mixing in the quark sector, we turn to 
a simple see-saw scenario~\cite{Mohapatra:1980yp} with Majorana neutrinos.
We introduce right-handed gauge singlet neutrino states, endowed with
a Majorana mass term $m_{R_i}$ and a regular Dirac mass term $m_{D_i}$
connecting left- and right-handed neutrino fields of the $i$-th generation.
For simplicity, we assume one right-handed neutrino per matter 
field generation and use $\nu_i$ and $N_i$ to denote the light and heavy 
neutrino states, respectively. 
The masses of the neutrino fields $\nu_i$ and $N_i$ are the eigenvalues 
of the see-saw neutrino mass matrix: 
\beq
	m_{\nu_i} = \frac{m_{D_i}^2}{m_{R_i}} + \ldots \quad \mathrm{and} \quad m_{N_i} = m_{R_i} + \ldots,
\eeq
where the ellipsis denotes terms of higher order in $m_{D_i}/m_{R_i}$ 
which are practically negligible.
In order to connect to the previous case, we allow the see-saw
mechanism only to mix neutrino states within one generation, i.e.\
we require that the six by six neutrino mass matrix has a threefold
block diagonal structure.
The couplings of the light and heavy Majorana neutrinos,
$\nu_i$ and $N_i$, to the $W$-boson and the charged lepton $l_j$ 
to lowest non-trivial order in $m_{\nu_i}/m_{N_i}$ 
are then given as:
\beq
	\mathcal{L}_\mathrm{int} = \frac{e}{\sqrt{2} s_W} V_{\nu_i l_j} \left(
		\sqrt{1-\frac{m_{\nu_i}}{m_{N_i}}} \; \bar \nu_i \gamma^\mu L l_j + 
		\sqrt{\frac{m_{\nu_i}}{m_{N_i}}} \;
		\bar N_i \gamma^\mu L l_j
	\right)	W_\mu.	
	\label{eq:intlagrange}
\eeq
In this ansatz, described in more detail in Ref.~\cite{Diener:2001qt}, 
LFV is entirely incorporated in
the matrix $V$ and factorises from the see-saw-induced, 
intra-generational mixing between the heavy and light neutrinos of one generation.

For light neutrinos, we have already derived the contribution to the transition
amplitude \mbox{$\langle \mu | j_{\mathrm{em}}^\mu (0) | e \rangle$} in the previous section.
In a see-saw scenario, this result still holds for the light degrees of freedom, 
except that the couplings have to be slightly modified to account for the
additional factor \mbox{$(1-m_{\nu_i}/m_{N_i})$}. 

Of course, an expansion of the formfactors $F_i$ 
in the ratio \mbox{$\epsilon = m_N^2/m_W^2$} as in 
section~\ref{sec:Dirac} makes no sense here, 
because we will in general have \mbox{$\epsilon > 1$}.

As in the previous section, we iteratively apply Eq.~(\ref{eq:propdec})
to rewrite the loop integrals and expand the form factors
in powers of squares of external momenta over 
powers of linear combinations of the $W$-boson and heavy 
neutrino mass squared.
At lowest non-trivial order in this expansion, the 
contribution from heavy Majorana neutrinos to the formfactors
introduced in section~\ref{sec:amp} can be written as 
\beqar
	F_1 &=& \sum_{j=1}^3 \frac{e g^2 V_{\nu_j e} V_{\nu_j \mu}^\star}{64 \pi^2 (m_{N_j}^2-m_W^2)^3} 
		\frac{m_{\nu_j}}{m_{N_j}}
		g(m_{N_j}^2, m_W^2) q^2, \\
	F_2 &=& \sum_{j=1}^3 \frac{- e g^2 V_{\nu_j e} V_{\nu_j \mu}^\star}{64 \pi^2 (m_{N_j}^2-m_W^2)^3} 
		\frac{m_{\nu_j}}{m_{N_j}}
		g(m_{N_j}^2, m_W^2) q^2, \\
%
	F_3 &=& \sum_{j=1}^3  \frac{e g^2 V_{\nu_j e} V_{\nu_j \mu}^\star}{256 \pi^2 (m_{N_j}^2-m_W^2)^3} 
		\frac{m_{\nu_j}}{m_{N_j}}
		g(m_{N_j}^2, m_W^2) \frac{m_{N_j}^2}{m_W^2} (m_e+m_\mu)^2 ,\\
	F_4 &=&\sum_{j=1}^3  \frac{- e g^2 V_{\nu_j e} V_{\nu_j \mu}^\star}{256 \pi^2 (m_{N_j}^2-m_W^2)^3}
	 		\frac{m_{\nu_j}}{m_{N_j}}
			g(m_{N_j}^2, m_W^2)  \frac{m_{N_j}^2}{m_W^2} (m_e-m_\mu)^2, 
\label{eq:Fheavynu}
\eeqar
where the function $g$ is given by
\beq
g(m_{N_j}^2, m_W^2) =  3 m_{N_j}^4 - 4 m_{N_j}^2 m_W^2 + m_W^4 - 2 m_{N_j}^4 
		\log \frac{m_{N_j}^2}{m_W^2}.
	\label{eq:g}
\eeq
Neglecting the electron mass, the transition amplitude, again, attains a particularly
simple form.
In this case, using the definitions of Eqs.~(\ref{eq:MpMm}) and~(\ref{eq:g}), we can express the heavy
Majorana neutrino contribution to the transition amplitude as
\beq
 \langle \mu | j_{\mathrm{em}}^\mu (0) | e \rangle = \sum_{j=1}^3
	\frac{e g^2 V_{\nu_j e} V_{\nu_j \mu}^\star}{32 \pi^2 (m_{N_j}^2-m_W^2)^3} 
		\frac{m_{\nu_j}}{m_{N_j}} g(m_{N_j}^2, m_W^2) 
	\left(q^2 \mathcal{M}^\mu_+ 
		+ \frac{1}{4} m_\mu^2 \frac{m_{N_j}^2}{m_W^2} \mathcal{M}^\mu_-  \right).
	\label{eq:majoranaamp}
\eeq

The assumption we have made about the structure of the leptonic mixing matrix,
namely that flavour mixing factorises from see-saw induced, intragenerational
mixing can easily be relaxed at this point.
To achieve this, one needs to replace the product of the mixing matrix 
elements and the neutrino mass ratio with two generalised mixing matrix
elements
\beq
	V_{\nu_j e} V_{\nu_j \mu}^\star \frac{m_{\nu_j}}{m_{N_j}} \to V_{N_j e} V_{N_j \mu}^\star.
\eeq
The structure and algebraic properties of such generalised mixing matrices are discussed 
in great detail in Refs.~\cite{Schechter:1980gr}.  

\section{The Cross Section \mbox{$e + \mathrm{N} \to \mu + \mathrm{N}$} at Low Energy}
\label{sec:Xsec}

In this section we consider the elastic electron to muon conversion cross
section in the coulomb field of a heavy nucleus with vanishing energy transfer.
The $S$-matrix element for this process is then given by
\beq
	S_{f i} = 2 \pi \delta (E_\mu - E_e) \frac{Z e}{q^2} F_\mathrm{N}(\vec q)  
	\langle \mu | j_{\mathrm{em}}^0 (0) | e \rangle ,
\eeq 
where $Z$ denotes the atomic number of the nucleus and $F_\mathrm{N}(\vec q)$ is 
the Fourier transform of the normalised nuclear charge distribution
\beq
	F_\mathrm{N}(\vec q) = \int \mathrm{d} \vec{x} e^{i \vec q \vec x} \rho (\vec x)
	\quad \mathrm{with} \quad \int \mathrm{d} \vec{x}\rho (\vec x) = 1.
\eeq
Under the assumption of a radially symmetric charge distribution, 
the differential cross section formula for electron to muon conversion
in the coulombic field of 
a heavy nucleus with vanishing energy transfer reads
\beq
	\frac{\mathrm{d} \sigma}{\mathrm{d} \Omega} = \frac{(Z e)^2}{(4 \pi)^2} \frac{1}{q^4} 
	|F_\mathrm{N}(q^2)|^2 |\langle \mu | j_{\mathrm{em}}^0 (0) | e \rangle|^2.
	\label{eq:dsigdomega}
\eeq
To evaluate the cross section, we need to calculate 
the square of the matrix element 
\mbox{$\langle \mu | j_{\mathrm{em}}^0 (0) | e \rangle$} averaged and summed over 
electron and muon spin, respectively. For vanishing energy transfer to the nucleus 
a straightforward calculation yields:
\beqar
\frac{1}{2} \sum_{\mathrm{spins}} \mathcal{M}_+^0 \left( \mathcal{M}_+^0 \right)^\star &=&
	\frac{1}{2} (q^2-m_\mu^2) + 2 E_e^2, \label{eq:squaremat1}\\
\frac{1}{2} \sum_{\mathrm{spins}} \mathcal{M}_-^0 \left( \mathcal{M}_-^0 \right)^\star &=&
	\frac{1}{2} (q^2-m_\mu^2) - 2 E_e^2 \frac{q^2}{m_\mu^2} , \\
\frac{1}{2} \sum_{\mathrm{spins}} \mathcal{M}_+^0 \left( \mathcal{M}_-^0 \right)^\star &=&
	\frac{1}{2} (q^2-m_\mu^2). \label{eq:squaremat3}
\eeqar

\subsection{The Cross Section for Three Light Dirac Neutrinos}
\label{sec:Xseclight}

Inserting the results Eqs.~(\ref{eq:squaremat1}--\ref{eq:squaremat3}) and 
Eqs.~(\ref{eq:diracff1}--\ref{eq:diracff4})
into the squared transition amplitude, we obtain
\beqar
    \lefteqn{|\langle \mu | j_{\mathrm{em}}^0 (0) | e \rangle|^2 = 
	\left( \frac{e g^2}{32 \pi^2} \right)^2 \left(
		V_{\nu_2 e} V^\star_{\nu_2 \mu} \epsilon_2  + 
		V_{\nu_3 e} V^\star_{\nu_3 \mu} \epsilon_3  \right)^2} \nonumber \\ 
	&& \times \left\{ \frac{q^4}{m_W^4} \left( \frac{1}{2} (q^2-m_\mu^2) + 2 E_e^2 \right) -
            \frac{1}{2} \frac{q^2 m_\mu^2}{m_W^4} \frac{1}{2} (q^2-m_\mu^2)  \right. \nonumber \\
	&& \left. 
	+  \frac{1}{16} \frac{m_\mu^4}{m_W^4} \left( \frac{1}{2} (q^2-m_\mu^2) 
	- 2 E_e^2 \frac{q^2}{m_\mu^2} \right) \right\}  .
\eeqar
For an approximately pointlike nuclear charge distribution, i.e.\ \mbox{$F_\mathrm{N}=1$}, 
the singly differential cross
section with respect to the cosine of the scattering angle reads:
\beqar
    \lefteqn{\frac{\mathrm{d} \sigma}{\mathrm{d} \cos \theta} = \frac{(Z e)^2}{8 \pi} \frac{1}{q^4}
		\left( \frac{e g^2}{32 \pi^2} \right)^2
	\left( V_{\nu_2 e} V^\star_{\nu_2 \mu} \epsilon_2  + 
	       V_{\nu_3 e} V^\star_{\nu_3 \mu} \epsilon_3  \right)^2 } \nonumber \\
	&& \times \frac{-m_\mu^6+9 m_\mu^4 q^2 + 16 q^4 (4 E_e^2+q^2) 
	- 4 m_\mu^2 q^2 (E_e^2+ 6 q^2)}{32 m_W^4}.
\label{eq:dsigdcth}
\eeqar 
In the limit of vanishing energy transfer to the nucleus and zero electron
mass, we obtain the following expression for the squared photon four-momentum
\beq
 	q^2 = m_\mu^2 - 2 E_e^2 \left(1 - \cos \theta \sqrt{1- m_\mu^2/E_e^2} \right),
\eeq
which allows us to express the cross section formula of Eq.~(\ref{eq:dsigdcth})
in terms of the electron energy and the scattering angle.

Clearly, $q^2$ never vanishes and the phase space integration  
can be performed analytically.
Defining \mbox{$x=m_\mu^2/E_e^2$}, one obtains for the total scattering cross section:
\beqar
 \lefteqn{ \sigma = \frac{(Z e)^2}{8 \pi}
		\left( \frac{e g^2}{32 \pi^2} \right)^2
	\left( V_{\nu_2 e} V^\star_{\nu_2 \mu} \epsilon_2  + 
	       V_{\nu_3 e} V^\star_{\nu_3 \mu} \epsilon_3  \right)^2 
	\frac{E_e^2}{64 m_W^4}} \nonumber \\ 
	&& \times  \left\{ 4 (32 - 9 x)  +
		 \frac{x (9 x-4 )}{\sqrt{1-x}}
	\log{\frac{8 (1 - x) + x^2 
	- 4 ( 2 - x) \sqrt{1-x}}{x^2}} \right\}.
	\label{eq:totalXS}
\eeqar

\subsection{The Cross Section Contribution from Heavy Majorana Neutrinos}
\label{sec:XsecMajorana}

With the help of Eqs.~(\ref{eq:squaremat1}--\ref{eq:squaremat3}), the square of the 
transition matrix element given in Eq.~(\ref{eq:majoranaamp}) of section~\ref{sec:Majorana}
can be written as
\beqar
	 \lefteqn{|\langle \mu | j_{\mathrm{em}}^0 (0) | e \rangle|^2 = 
	  \left(\frac{e g^2}{32 \pi^2}\right)^2 \left\{ q^4 
		\left(\frac{1}{2} (q^2-m_\mu^2) + 2 E_e^2\right) \left|\sum_{j=1}^3 G_j\right|^2 
		\right. }\nonumber  \\
	&& + \frac{1}{2} q^2 m_\mu^2 \frac{1}{2} (q^2-m_\mu^2) 
		\Re \left( \sum_{j=1}^3 G_j \right)
	  \left( \sum_{i=1}^3 G_i \frac{m_{N_i}^2}{m_W^2}\right)^\star  \nonumber \\
	&& \left. + \frac{1}{16} m_\mu^4  \left(\frac{1}{2} (q^2-m_\mu^2) - 2 E_e^2 \frac{q^2}{m_\mu^2}\right) 
	\left|\sum_{j=1}^3 G_j \frac{m_{N_j}^2}{m_W^2}\right|^2 \right\}, 
\label{eq:squareampmajo}
\eeqar
where we have made use of the abbreviation
\beq
	G_j = \frac{V_{\nu_j e} V_{\nu_j \mu}^\star}{(m_{N_j}^2-m_W^2)^3} 
		\frac{m_{\nu_j}}{m_{N_j}} g(m_{N_j}^2, m_W^2).
\eeq
For a pointlike nuclear charge distribution we can easily insert the result of 
Eq.~(\ref{eq:squareampmajo}) for 
the squared transition matrix element into the cross section 
formula Eq.~(\ref{eq:dsigdomega}) and integrate over solid angle:
\beqar
	\lefteqn{\sigma = \frac{(Z e)^2}{8 \pi}	\left(\frac{e g^2}{32 \pi^2}\right)^2 E_e^2
	  \left(2 \left|G \right|^2 + \frac{x}{2} \Re G \tilde G^\star 
		-\frac{x}{16} \left|\tilde G\right|^2 + \frac{x}{64 \sqrt{1-x}} \right. } \nonumber \\
	&&  \left.
  	  \times
	\left((x-4) \left|\tilde G \right|^2 
		- 8 x \Re  G \tilde G^\star \right) \log{\frac{8 (1 - x) + x^2 
	- 4 ( 2 - x) \sqrt{1-x}}{x^2}}  \right), 
\eeqar
where we have, again, substituted \mbox{$x = m_\mu^2/E_e^2$} and introduced the notation
\mbox{$G = \sum_{i=1}^3 G_i$} and \mbox{$\tilde G = \sum_{i=1}^3 G_i \frac{m_{N_i}^2}{m_W^2}$}.

\section{Numerical Results and Conclusions}
\label{sec:res}

We proceed with the numerical evaluation of the cross section 
formulae derived above.
Considering a two-neutrino oscillation scenario with mixing
angle $\theta_{12}$ and masses \mbox{$m_{\nu_1}^2 \approx 0$} and
\mbox{$\Delta m_\nu^2 \approx m_{\nu_2}^2$}, we can write 
\beq
    V_{\nu_2 e} V^\star_{\nu_2 \mu} \epsilon_2 = \frac{\Delta m_\nu^2}{m_W^2} \frac{1}{2} \sin 2 \theta_{12}.
\eeq
Using this result to evaluate Eq.~(\ref{eq:totalXS}) for $E_e = 2 m_\mu$, 
which reduces background from muon pair production, and $Z=74$
for a tungsten target,
we obtain:
\beq 
	\sigma \approx 6 \dot 10^{-50} \;  
		\left( \frac{\Delta m_\nu^2}{\mathrm{eV}^2} \right)^2 
		\frac{1}{4} \sin^2 2 \theta_{12} \, \mathrm{pb}. 
\label{eq:diracres}
\eeq 
Despite maximal mixing, a LFV signal is 
strongly suppressed because, firstly,
the expansion of the flavour off-diagonal amplitude 
in \mbox{$\epsilon_j = m_{\nu_j}^2/m_W^2$} starts at 
first order in $\epsilon_j$, and, secondly, 
the first non-vanishing contribution in this expansion, when reduced
to vacuum integrals, is of the order of \mbox{$k_\mathrm{ext}^2/m_W^2$}.
These effects combined result in a factor
\mbox{$k_\mathrm{ext}^4 (\Delta m_\nu^2)^2/m_W^{8}$} in the squared 
amplitude.\\

In case of the see-saw scenario outlined in section~\ref{sec:Majorana}, 
Eq.~(\ref{eq:diracres}) still holds
for the light degrees of freedom as the extra factor
\mbox{$\sqrt{1-m_\nu^2/m_N^2}$} of Eq.~(\ref{eq:intlagrange}) can be neglected
for all practical purposes. 

Considering the contribution from heavy Majorana neutrinos outlined in
section~\ref{sec:Majorana}, the structure of the formfactors $F_i$
suggests, that a sizeable effect, if at all, can be expected if 
the heavy neutral lepton states have masses only little above the
present $95\%$ confidence 
limit~\cite{Hagiwara:fs} of $80.5\,\mathrm{GeV}$.

To compare the contributions from the heavy Majorana neutrinos
with the previous case of light Dirac neutrinos, we confine ourselves
to two mixing generations and 
choose a universal heavy Majorana neutrino mass of \mbox{$m_N= 2 m_W$}.
This entails the following simplification:
\beq
	G_1 + G_2 = \frac{1}{2} \sin 2 \theta_{1 2} 
	\frac{33-32 \log 4}{54 m_W^2} \frac{\Delta m_\nu}{m_W},
\eeq
which, for $E_e$ and $Z$ as above, leads to 
\beq
	\sigma \approx 3 \cdot 10^{-29}
	\left(\frac{\Delta m_\nu}{\mathrm{eV}}\right)^2 \frac{1}{4} \sin^2 2 \theta_{12} \, \mathrm{pb},
	\label{eq:Majores}
\eeq
which amounts to an ehancement by more than twenty orders of magnitude compared to the 
minimal Dirac case.

The practical relevance of the numbers given in Eqs.~(\ref{eq:diracres}) 
and~(\ref{eq:Majores}) can be evaluated by giving a rough order 
of magnitude estimate of the target luminosity of 
a $1\,\mathrm{mA}$ electron beam on a tungsten target.
Assuming as an approximation that all electrons interact within a depth of one radiation length 
$X_0$ in the target, the luminosity $\mathcal{L}$ can be calculated as
\beq
	\mathcal{L} \approx \frac{X_0}{A_\mathrm{W}} \frac{1\,\mathrm{mA}}{e} \approx 0.1\,
	\mathrm{fb}^{-1}/\mathrm{s},
\eeq
where $A_\mathrm{W}=183.84\, \mathrm{u}$ and \mbox{$X_0 = 6.76\,\mathrm{g}/\mathrm{cm}^2$} 
are the nuclear mass number in atomic mass units
and the electromagnetic radiation length of tungsten 
and $e$ denotes the electron charge magnitude. \\

This order of magnitude estimate shows that even in an optimistic see-saw scenario 
with low-lying heavy neutrino states, where
the cross section is enhanced by as much as twenty orders of magnitude compared to the 
minimal Dirac case, a LFV signal is far out of reach.
We must conclude that an electron-nucleus scattering experiment at low electron 
energies will not be able to find electron to muon conversion if
LFV arises from conventional SM lepton mixing as outlined above.

\section*{Acknowledgments}

I would like to thank Ansgar Denner and Michael Spira 
for useful discussions and careful proofreading of the manuscript. 
Andr\'e Sch\"oning is gratefully acknowledged for pointing out this interesting
subject and for useful discussions as well as proofreading of the manuscript.

\end{document}